\documentclass[twocolumn]{aastex62}

\begin{document}
\title{Phosphorus Abundances in the Hyades and Galactic Disk}

\author{Z. G. Maas}
\affil{Indiana University Bloomington, Astronomy Department, 727 East Third Street, Bloomington, IN 47405, 
USA}
\author{G. Cescutti}
\affil{INAF, Osservatorio Astronomico di Trieste, Via Tiepolo 11, I-34143 Trieste, Italy}
\affil{IFPU - Institute for Fundamental Physics of the Universe, via Beirut 2, 34151, Trieste, Italy}
\author{C. A. Pilachowski}
\affil{Indiana University Bloomington, Astronomy Department, 727 East Third Street, Bloomington, IN 47405, 
USA}

\email{zmaas@indiana.edu}

\begin{abstract}

We have measured phosphorus abundances in nine disk stars between -1 $<$ [Fe/H] $<$ --0.5 and in 12 members of the Hyades open cluster using two P I lines at 1.06 $\mu$m. High resolution infrared spectra were obtained using Phoenix on Gemini South and abundances were determined by comparing synthetic spectra to the observations. The average abundance for the dwarf stars in our Hyades sample was $<$ [P/Fe] $>$ = --0.01 $\pm$ 0.06 and $<$ [P/Fe] $>$ = 0.03 $\pm$ 0.03 dex for the three giants. The consistency suggests abundances derived using the 1.06 $\mu$m P I lines are not subjected to temperature or luminosity dependent systematic effects at high metallicities. Our [P/Fe] ratios measured in disk stars are consistent with chemical evolution models with P yields increased by a factor of 2.75. We find  [P/O], [P/Mg], [P/Si] and [P/Ti] ratios are consistent with the solar ratio over a range of --1.0 $<$ [Fe/H] $<$ 0.2 with the [P/Si] ratio increasing by $\sim$ 0.1 - 0.2 dex at the lowest [Fe/H] ratios. Finally, the evolution of [P/Fe] with age is similar to other $\alpha$ elements, providing evidence that P is produced at the same sites. 

\end{abstract}


\section{Introduction}
\label{sec::intro}

Phosphorus is an important element for life and the seventeenth most abundance element in the Universe. P has one stable isotope ($^{31}$P) and is thought to be primarily produced from neutron capture on Si isotopes in hydrostatic carbon and neon burning in massive stars \citep{woosley95}. Type Ia SN may also produce P \citep{travaglio04,leung18} but are thought to be less significant source than burning in massive stars. Finally, little to no phosphorus is thought to be produced in AGB stars \citep{karakas16}. 

Abundance measurements of the odd-Z element phosphorus in stars have grown significantly in number recently and have begun to constrain chemical evolution models of P in the Galaxy. The first chemical evolution study of P I used near-IR features at 1.06 $\mu$m and measured P in $\sim$ 20 FGK dwarfs \citep{caffau11}. Additional abundances measurements in metal poor stars (--4 $<$ [Fe/H] $<$ --0.2) were performed using P I lines in the UV \citep{roederer14,jacobson14}. Comparisons to chemical evolution models for these initial studies found P is under-predicted by models and yields need to be increased by a factor of 2.75 to match the data \citep{cescutti12,jacobson14}. Additionally, the inclusion of hypernova production of P is necessary to match [P/Fe] ratios measured in metal poor stars  \citep{cescutti12,jacobson14}.

%Multiple additional studies have begun to explore P abundances in the Galaxy.

Targeted surveys with the Y-band P I features have confirmed an offset between chemical evolution models and abundance measurements over the --0.6 $<$ [Fe/H] $<$ 0.2 metallicity range \citep{caffau16,maas17,caffau19}. Additional P abundances have also been measured in individual metal poor stars \citep{spite17} further constraining P production in the early Universe. The APOGEE survey has also measured P \citep{holtzman18}, however the P I features in the H-band are weak, blended with telluric features, and blended with other atomic/molecular features. As a result abundance determinations are difficult and there is significant scatter in measured P abundances from H band features. \citep{jonsson18,holtzman18}.

P abundances from APOGEE spectra show that the P abundance increases as metallicity decreases \citep{hawkins16}. APOGEE data also finds [P/Fe] ratios in Sagittarius dwarf galaxy stars are lower than disk stars and are consistent with nucleosynthesis in core collapse supernova (CCSNe) \citep{hasselquist17}. Finally, a difference in [P/Mg] ratios in high [Mg/Fe] and low [Mg/Fe] populations using APOGEE data suggests P may be produced in Type Ia supernovae \citep{weinberg19}. In addition to APOGEE, P has also been measured in H-band IGRINS spectra in horizontal branch stars \citep{afsar18}. Finally, a mean P I abundance of 0.10 $\pm$ 0.12 dex was measured in the open cluster NGC 6940 \citep{topcu19}. 

To further constrain the nucleosynthesis of P, we have measured abundances in 9 stars between --1 $<$ [Fe/H] $<$ --0.5 and 12 stars in the Hyades open cluster using the P I lines at 1.06 $\mu$m in stars. By selecting a targeted sample of metal poor stars we can compare precise P abundances to chemical evolution models. Comparisons of [P/Fe] and P to various $\alpha$ elements will constrain theoretical predictions of how P is produced. Also, the Hyades open cluster sample provides an opportunity to measure P abundances at the high metallicity range ($\sim$ [Fe/H] = 0.2 dex) at an age of 625 Myr \citep{perryman98}, and to test for systematic effects on abundance as a function of T$_{\mathrm{eff}}$. 

Details of our sample selection and observations are given in section \ref{sec::obs}. Our abundance determination methodology and uncertainty calculations are explained in \ref{sec::method}. The discussion of our results and in the context of chemical evolution and nucleosynthesis models is located in section \ref{sec::discussion}. Finally, our conclusions are summarized in section \ref{sec::conclusions}. 

\section{Observations and Data Reduction}
\label{sec::obs}

Observations were performed using Phoenix \citep{hinkle98} on Gemini South as apart of proposal GS-2017B-Q-47. Phoenix is a high resolution infrared spectrometer capable of observing between 1$\mu$m - 5$\mu$m. We chose the 4 pixel slit width to achieve a resolution of R $\sim$ 50000. To observe the P I features at 10581 $\mbox{\AA}$ and 10596 $\mbox{\AA}$, we chose a central wavelength of 10593 $\mbox{\AA}$, with a total range of 50 $\mbox{\AA}$. We only selected stars with known atmospheric parameters since the spectral region covered by Phoenix contains too few absorption lines to determine atmospheric parameters independently. We also required sufficiently bright stars (J $\lesssim$ 7.0 mag) to achieve the signal to noise needed to measure weak P I features in our relatively metal poor stars. We selected our sample stars to measure stars below [Fe/H] $<$ --0.5 where the few existing [P/Fe] abundances measurements are enhanced relative to the solar ratio. The P I features at 1.06 $\mu$m however become undetectable at [Fe/H] $\lesssim$ --1. We therefore chose stars over the metallicity range  --1 $<$ [Fe/H] $<$ --0.5 from the sample of \citet{bensby14}, which include both atmospheric parameters and abundance measurements for other elements that would be useful to compare to P. Nine stars in total were observed and the observations are listed in Table \ref{table::obslog}.

We also included Hyades dwarf and giant stars with homogeneous abundances to probe potential systematic difference in [P/Fe] between low temperature giants and dwarf stars using the P I features at 1.06 $\mu \mathrm{m}$. We chose the Hyades because several dwarf stars are available at J $\lesssim$ 7.0 mag at a range of atmospheric parameters, three non-binary giants exist in the cluster, and its young age and high metallicity ([Fe/H] = 0.11 \citealt{takeda13}) provide constraints on the younger, more metal rich stellar population. Spectra were obtained for nine dwarf stars with temperatures ranging from 5735 $<$ T$_{\mathrm{eff}}$ $<$ 6291 K and three giants as shown in Table \ref{table::obslog}.

\begin{deluxetable}{ c c c c c}
%\tabletypesize{10pt} 
%\rotate
\tablewidth{0pt} 
\tabletypesize{\footnotesize}
\tablecaption{Summary of Phoenix Observations \label{table::obslog}} 
 \tablehead{\colhead{Identifier} & \colhead{UT Date}& \colhead{J\tablenotemark{a}} & \colhead{Spectral\tablenotemark{b}} & \colhead{S/N} \\
 \colhead{} & \colhead{} & \colhead{(Mag.)} & \colhead{Type} & \colhead{} }
\startdata
\hline
\multicolumn{5}{c}{Field Stars} \\
\hline
HIP 6949   & 2017 Dec 1 & 6.133  & G8IV/V & 210\\
HIP 7961   & 2017 Dec 4  &  6.233	&	G2/3V & 220\\
HIP 10798   & 2017 Dec 3  &  5.056	&	G8V & 200\\
HIP 14086   & 2017 Dec 3  &  4.364	&	K2V & 230\\
HIP 17147  & 2017 Dec 1 & 5.588 & F7/8V & 330 \\
HIP 30439   & 2017 Dec 3  &  6.491	&	G6V & 260\\
HIP 33324   & 2017 Dec 4  &  6.231	&	G2V & 230\\
HIP 38625   & 2017 Dec 3  &  5.972	&	K0V & 70\\
HIP 44075   & 2017 Dec 4  &  4.710	&	G2V & 230\\
\hline
\multicolumn{5}{c}{Hyades Cluster Members} \\
\hline
HIP 15304 & 2017 Dec 1   & 6.291 &   F8 & 160 \\
HIP 15310  & 2017 Dec 4  &  6.639	&	G0 & 130\\
HIP 19793 &  2017 Dec 3 &   6.877 &  G5V & 140\\
HIP 19796 & 2017 Dec 3  &  6.128	& F8V & 280\\
HIP 20205  & 2017 Dec 4  &  1.300	&	G9.5III & 80\\
HIP 20237  & 2017 Dec 4  &  6.395	&	G0V &180\\
HIP 20455   & 2017 Dec 4  &  1.300	&	G9.5III & 90\\
HIP 20741 &  2017 Dec 1& 6.893   &   G2V &  110\\
HIP 20889   & 2017 Dec 4  &  1.350	&	G9.5III & 90\\
HIP 20899 & 2017 Dec 4  &  6.696	&	G2V &160\\
HIP 21112  & 2017 Dec 4  &  6.733	&	F9V &120\\
HIP 22422  & 2017 Dec 4  &  6.642	& F8	 &110\\
\enddata
\tablenotetext{a}{J magnitudes from 2MASS \citep{skrutskie06}}
\tablenotetext{b}{Spectral types from the SIMBAD database}
\end{deluxetable}

\begin{figure}[htp]
	\centering 
 	\includegraphics[trim=0cm 0cm 0cm 0cm, scale=.44]{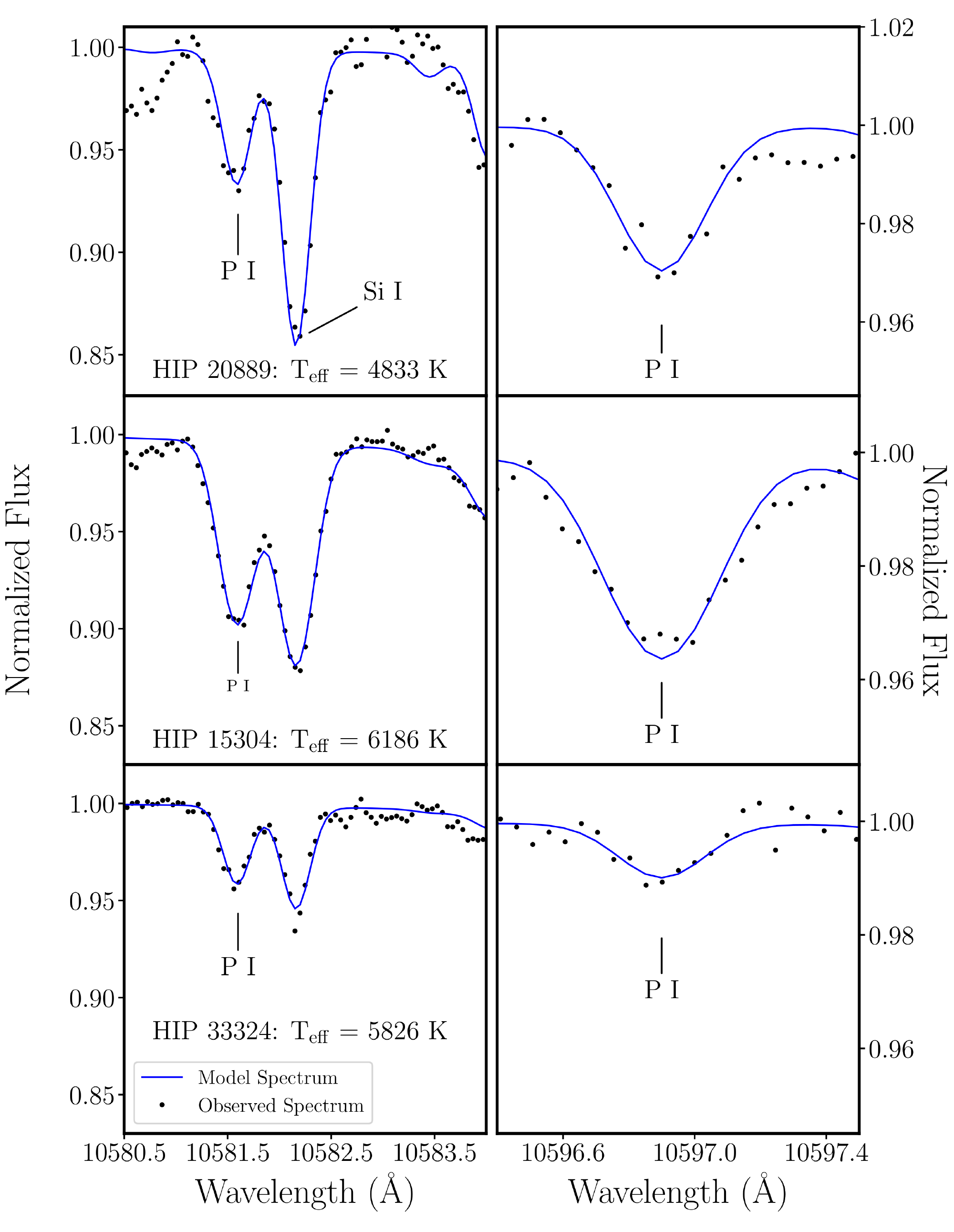}
	\caption{Spectra and synthetic spectra fits to the  Hyades giant HIP 20889 (top panel), the Hyades dwarf HIP 15304 (middle panel), and the disk star HIP 33324 (bottom panel). The left column shows fits to the P feature at 10581.5 $\mbox{\AA}$ while the right column shows the fit to the weaker P I feature at 10596.9 $\mbox{\AA}$. \label{fig:spectra} }
	\end{figure}

Data reduction utilized the IRAF software suite\footnote{IRAF is distributed by the National Optical Astronomy Observatory, which is operated by the Association of Universities for Research in Astronomy, Inc., under cooperative agreement with the National Science Foundation.} and followed the same procedures described in \citet{maas17}. To summarize the reduction steps, the images were trimmed, the flat images were dark-subtracted, and flats were median combined. When taking exposures, the objects were nodded along the slit in an 'abba' pattern and then subtracted from one another to remove telluric contamination. The object exposures were flatfielded and extracted, and a wavelength solution was applied to each spectrum. Due to the limited wavelength window, no telluric lines and extremely few standard emission lamp lines were available; the wavelength solution was derived using stellar lines in the region. Finally, the spectra were averaged and normalized. Examples of the two phosphorus features in the final, reduced spectra are shown in Figure \ref{fig:spectra}.  

\section{Stellar Abundance Measurements}
\label{sec::method}
\subsection{Abundance Measurement Methodology}
\label{subsec::abundance_measurement_methodology}

As stated in section \ref{sec::obs}, atmospheric parameters for our target stars have been derived in previous studies. The disk star sample was studied by \citet{bensby14}, Hyades dwarfs' atmospheric parameters were obtained from \citet{takeda13}, and finally Hyades giants were obtained from \citet{dutra16}. Atmospheric parameters from \citet{takeda13} were adopted because they had self-consistently derived temperature, log(g), metallicity, microturbulence, and vsini. The atmospheric parameters for these stars are listed in Table \ref{table::results}. 

The effective temperatures adopted from the literature and other atmospheric parameters were computed using different methodologies. First, \citet{bensby14} derived their atmospheric parameters using equivalent widths from Fe I and Fe II lines. Differences between effective temperatures and surface gravities derived from stars with accurate parallax values were removed using an empirical correction. The analysis from \citet{takeda13} determined their effective temperatures and log(g) from the known luminosities and masses of the Hyades stars; derived in \citet{bruijne01}. Finally, \citet{dutra16} compared temperatures from spectroscopic methods for the entire sample of Hyades stars, the infrared flux method for dwarf stars, and from angular diameters for the giant stars. Using a line-list with few blends and excluding lines with potential NLTE effects, they were able to determine consistent atmospheric parameters and [Fe/H] abundances for both the dwarf and giant stars. 

To test that the different atmospheric parameters are consistent, we derive atmospheric parameters for the stars in our sample using 2MASS photometry \citep{skrutskie06}, V band photometry adopted from the SIMBAD database, and the T$_{\mathrm{eff}}$-color relation from \citet{gonzalez09}. We removed stars with J and K$_{s}$ uncertainties above 0.20 mags and converted the K magnitudes of the three Hyades giants to the K$_{s}$ photometric band. We find the average difference between literature effective temperatures of our sample and J - K$_{s}$ derived effective temperatures is --32 $\pm$ 126 K. The V - K$_{s}$ required a correction of 0.03 mag to remove a systematic offset between the photometric and the literature values. Without the offset, the average difference is --31 $\pm$ 70 K and with the offset a values of 9 $\pm$ 64 K. In either case, the standard deviation is consistent with literature T$_{\mathrm{eff}}$ measurement errors, indicating both Hyades and disk star effective temperatures are compatible. The constancy of effective temperatures ensures no systematic offset is likely due to different atmospheric parameter derivation methodologies. We do not determine full independent atmospheric parameters because too few Fe lines exist in our spectral region to independently determine [Fe/H], microturbulence, or vsini.

Abundance measurements were carried out using MOOG (\citealt{sneden73}; v. 2017) with MARCS model atmospheres \citep{gustafsson08}. Plane-parallel atmospheric models were used for dwarf stars while spherical models were employed for the three Hyades giants in our sample. We used the  phosphorus atomic data from \citet{berzinsh97}, which has been used to determine abundances in \citet{caffau07,caffau11,caffau16,maas17,caffau19}. The \texttt{emcee} Python package was also used to achieve the best fit between the model and the data \citep{foreman13}. A log-likelihood function, shown in Eq. \ref{equation:loglikelihood}, was used to find the best fitting synthetic spectrum to the observed absorption features.  

\begin{equation}\label{equation:loglikelihood}
\footnotesize
\mathrm{log}(L) = -0.5 \sum \frac{D - \mathrm{synth}(\lambda,\mathrm{A(P)},\mathrm{A(Si)})}{(\sigma*f)^2.}  + \mathrm{log}\left(\frac{1}{(\sigma*f)^2} \right) \, ,
\end{equation}

\noindent In Equation~\ref{equation:loglikelihood}, synth is the synthetic spectrum created by MOOG and $D$ represents the observed data. Three parameters are varied when producing the synthetic spectrum; a wavelength shift of $\lambda$, and the input abundances of A(P) and A(Si). Finally, $f$ represents deviations from the expected signal to noise of 100 (initial $\sigma$ was assumed to be 0.01). For each iteration, a new synthetic spectrum was generated with parameters for $\lambda$, A(P), and A(Si). The 14$\%$ and 86$\%$ values from the posterior distribution were reported as the uncertainty on the fits. The final best fit to the data using this method for three example stars is shown in Fig. \ref{fig:spectra}. 

The Markov Chain Monte Carlo (MCMC) fitting methodology was tested by repeating abundance measurements for the P I and Si I features at 10581.57 $\mbox{\AA}$ and 10582.16 $\mbox{\AA}$ on the stars with known P abundance measurements from \citet{maas17}. Abundances were re-derived using the MCMC code and likelihood equation in Eq. \ref{equation:loglikelihood} then compared to the previous $\chi^{2}$ minimization results. We found an average difference of [Si/Fe] = 0.00 $\pm$ 0.05 dex for Si abundance measurements and [P/Fe] = 0.00 $\pm$ 0.08 dex in the 19 stars with Phoenix spectra. One star had an anomalous difference of --0.23 dex (the MCMC abundance -- \citet{maas17} abundance) which may be due to the low signal-to-noise of the spectrum for this object (S/N $\sim$ 70). HD 163363 was also indicated as an outlier when comparing abundance fits by eye to $\chi^{2}$ minimization methods in \citet{maas17} and without this star the average difference becomes [P/Fe] = 0.01 $\pm$ 0.06 dex for phosphorus. The uncertainties associated with the fits to Si and P are consistent with the standard deviation of the average differences found when comparing the samples. For example, the difference of [P/Fe] = 0.00 $\pm$ 0.08 for P is similar to the uncertainty on the fit of 0.05 - 0.07 dex for the MCMC fits in Table \ref{table::uncertainties} and an average uncertainty on the fit of 0.05 dex from \citet{maas17}.

\begin{deluxetable*}{ c c c c c c c c c c c c c }
%\tabletypesize{10pt} 
%\rotate
\tablewidth{0pt} 
\tabletypesize{\footnotesize}
\tablecaption{Abundance Results \label{table::results}} 
 \tablehead{\colhead{Identifier} & \colhead{T$_\mathrm{{eff}}$ }& \colhead{log g} & \colhead{[Fe/H]} & \colhead{$\xi$} & \colhead{vsini} & \colhead{ref.} & \colhead{[Si/Fe]} & \colhead{$\delta$[Si/Fe]} & \colhead{[P/Fe]} & \colhead{$\delta$[P/Fe]} & \colhead{[P/Fe]} & \colhead{$\delta$[P/Fe]} \\
 \colhead{} & \colhead{(K)} & \colhead{} & \colhead{} & \colhead{(km s$^{-1}$)} & \colhead{(km s$^{-1}$)} & \colhead{} & \colhead{} & \colhead{}  & \colhead{10581 $\mbox{\AA}$} & \colhead{10581 $\mbox{\AA}$} & \colhead{10596 $\mbox{\AA}$} & \colhead{10596 $\mbox{\AA}$}  }
\startdata
\hline
\multicolumn{5}{c}{Field Stars} \\
\hline
HIP 6949   & 5005 & 3.35  & --0.50 & 0.96 & \nodata & 1 & 0.20 & 0.03 & 0.29 & 0.06 & 0.38 & 0.12 \\
HIP 7961   & 5648  &  3.91	&	--0.62 & 1.08 & \nodata & 1 & 0.30 & 0.03 & 0.42 & 0.06 & 0.33 & 0.10 \\
HIP 10798   & 5302  &  4.61	& --0.51	& 0.65 &\nodata & 1 & 0.12 & 0.05 & 0.31 & 0.09 & 0.22 & 0.17 \\
HIP 14086   & 5095  &  3.47	&	--0.56 & 1.03 & \nodata & 1 & 0.18 & 0.03 & 0.18 & 0.08 & 0.14 & 0.08\\
HIP 17147  & 5970 & 4.52 & --0.81 & 1.27 & \nodata & 1 & 0.29 & 0.03 & 0.49 & 0.08 & 0.47 & 0.10\\
HIP 30439   &  5353  &  3.79	&	--0.57 & 0.85 & \nodata & 1 & 0.19 & 0.06 & 0.31 & 0.03 & 0.31 & 0.09 \\
HIP 33324   & 5826  &  4.24	&	--0.68 & 1.12 & \nodata & 1 & 0.27 & 0.03 & 0.28 & 0.04 & 0.19 & 0.06\\
HIP 38625   & 5188  &  4.40	&	--0.91& 0.62 & \nodata & 1 & 0.23 & 0.13 & 0.52 & 0.22 & \nodata & \nodata \\
HIP 44075   & 5937  &  4.22	&	--0.90 & 1.32 & \nodata & 1 & 0.29 & 0.03 & 0.50 & 0.05 & 0.63 & 0.06\\
\hline
\multicolumn{5}{c}{Hyades Cluster Members} \\
\hline
HIP 15304 & 6079   & 4.45 &   0.26 & 1.3  & 5.5 & 2 & 0.01 & 0.04 & 0.05 & 0.07 & 0.03 & 0.06\\
HIP 15310 & 5894  &  4.48	&	0.25 & 1.1 & 6 & 2 & 0.05 & 0.03  & 0.05 & 0.08  & 0.10 & 0.18\\
HIP 19793 &  5757 &   4.51 &  0.16 & 1.0& 5.5 & 2 & --0.06 & 0.04 & --0.12 & 0.08 & 0.04 & 0.09\\
HIP 19796  & 6291  &  4.40	& 0.14 & 1.5 & 13.4 & 2 & --0.05 & 0.06 & --0.06 & 0.08 & --0.05 & 0.09\\
HIP 20205   & 4875  &  2.71	&	0.11 & 1.43 & \nodata & 3 & 0.03 & 0.06 & 0.04 & 0.10 & 0.01 & 0.09\\
HIP 20237 & 6107  &  4.44	&	0.19 & 1.3  & 8.9 & 2 & --0.07 & 0.04 & --0.08 & 0.08 & --0.06 & 0.05\\
HIP 20455  & 4816  &  2.55	&	0.08 & 1.35 & \nodata & 3 & 0.04 & 0.05 & 0.05 & 0.09 & 0.18 & 0.09\\
HIP 20741 &  5735& 4.51   &   0.19 &  1.0 & 4.3 & 2 & --0.05 & 0.03 & --0.01 & 0.09 & 0.07 & 0.08\\
HIP 20889   & 4833  &  2.74	&	0.17 & 1.41& \nodata & 3 & --0.01 & 0.06 & --0.01 & 0.10 & 0.06 & 0.08\\
HIP 20899  & 5924  &  4.47	&	0.11 & 1.1  & 6 & 2 &0.01 & 0.03 & 0.00 & 0.07 & 0.25 & 0.10\\
HIP 21112 & 6186  &  4.42	&	0.14 & 1.4 & 4.9 & 2 & 0.06 & 0.04 & --0.02 & 0.06 & 0.06 & 0.06\\
HIP 22422 & 6037  &  4.45	& 0.16	 & 1.2  & 5.2 & 2 & 0.03 & 0.04 & 0.06 & 0.08 & 0.00 & 0.06\\
\enddata
\tablecomments{Sources: (1) \citealt{bensby14}; (2) \citealt{takeda13}; (3) \citealt{dutra16} \\ 
               Adopted A(X)$|_{\odot}$: A(Si) = 7.51 \citep{asplund09}, A(P) = 5.46 \citep{caffau07}, A(Fe) = 7.50 \citep{asplund09}}

\end{deluxetable*}

\subsection{Uncertainties}
\label{subsection::uncertainties}
The uncertainty on each abundance measurement due to noise in the spectra and uncertainty in the atmospheric parameters was calculated, assumed to be independent, added in quadrature, and listed in Table \ref{table::results}. The uncertainty in each abundance measurement due to uncertainty in each atmospheric parameter was derived by using new atmospheric models varied by 1$\sigma$ for a particular atmospheric parameter, then new abundances were derived by fitting the new synthetic spectra to the observed spectra. For each model, one parameter was varied 1$\sigma$ (e.g. T$_{\mathrm{eff}}$) while the others were held constant to the values listed in Table \ref{table::results}. 

A grid of synthetic spectra were derived in steps of either 0.01 dex in P or Si abundance, and the best fitting abundances were found by minimizing the $\chi^{2}$ between the model fit and the synthetic spectra. The 1$\sigma$ uncertainties on the atmospheric parameters were adopted from atmospheric parameter sources of \citet{takeda13,bensby14,dutra16} with individual star reference assignments listed in Table \ref{table::results}. As stated in subsection \ref{subsec::abundance_measurement_methodology}, the uncertainties on the fit were the 14$\%$ and 86$\%$ values derived from the MCMC fitting methodology. Each uncertainty was added in quadrature and the average uncertainties for each star are included with the results shown in Table \ref{table::uncertainties}.

\begin{deluxetable}{ c c c c }
%\tabletypesize{10pt} 
%\rotate
\tablewidth{0pt} 
\tabletypesize{\footnotesize}
\tablecaption{P Uncertainty Averages \label{table::uncertainties}} 
 \tablehead{\colhead{Uncertainty Source} & \colhead{Disk Stars}& \colhead{Hyades Giants} & \colhead{Hyades Dwarfs} \\ \colhead{} & \colhead{(dex)} & \colhead{(dex)} & \colhead{(dex)}}
\startdata
T$_{\mathrm{eff}}$   & 0.04 &  0.06 & 0.04 \\
log(g)  & 0.03 & 0.04 & 0.02 \\
$\mathrm{[Fe/H]}$ & 0.01 & 0.01 & 0.01 \\
$\xi$ & 0.01 & 0.01 & 0.01 \\
Fit & 0.06 & 0.07 & 0.05 \\
N in Sample & 9 & 9 & 3 \\
\enddata

\end{deluxetable}

\subsection{Examining Systematic Uncertainties in P Abundances}

\subsubsection{P Abundance Comparison between both P I Features}

Our Hyades sample provides a set of chemically homogeneous stars suitable to test if both P I lines give consistent P abundances and if significant systematic effects are present with our 1D LTE methodology. First, we compare the abundances of the 10581 $\mbox{\AA}$ and the 10596 $\mbox{\AA}$ P I features. Differences in abundance between the lines may indicate an unknown systematic effect on one or both lines, such as blends with other weak absorption features. Abundances from both P I lines in each star in our sample are compared in Fig. \ref{fig:line_comp}. We find the average difference between between the two P I lines (in the order 10581 $\mbox{\AA}$ - 10596 $\mbox{\AA}$) is $\Delta$[P/Fe] = --0.03 $\pm$ 0.09 dex. The average uncertainty for the 10581 $\mbox{\AA}$ feature for each stellar population is given in Table \ref{table::uncertainties} at a typical value of 0.06 dex and the average value for the 10596 $\mbox{\AA}$ feature is 0.06 dex. Added in quadrature results in a fit uncertainty of 0.08 dex, consistent with the standard deviation of the average differences of the abundance determinations from the two features. Since two lines are not sufficient to do a statistical analysis (e.g. Fe I lines) and averaging the two lines will not lower systematic uncertainties (e.g. from the atmospheric parameters) the results discussed further will use the 10581 $\mbox{\AA}$ abundances, as these lines are stronger with lower or comparable uncertainties; especially for the disk sample of stars observed which have weak 10596 $\mbox{\AA}$ features (as seen in Fig. \ref{fig:spectra}).

\subsubsection{P Abundances vs. T$_{\mathrm{eff}}$}

The P I lines used in the abundance analysis have high excitation potentials, slightly below 7 ev \citep{berzinsh97}. Although the excitation potential is high, the atomic transitions for these lines are similar to S I lines that show no strong NLTE effects \citep{asplund09}. No study has been performed to test for potential NLTE effects on phosphorus abundance derivations using the P I lines. To quantify the potential NLTE effects on measured P abundance empirically, we examine if a relationship exists between effective temperature (and luminosity between dwarfs and giants) and P abundance measurements in our Hyades sample of stars. 

\begin{figure}[h!]
	\centering 
 	\includegraphics[trim=0cm 0cm 0cm 0cm, scale=.47]{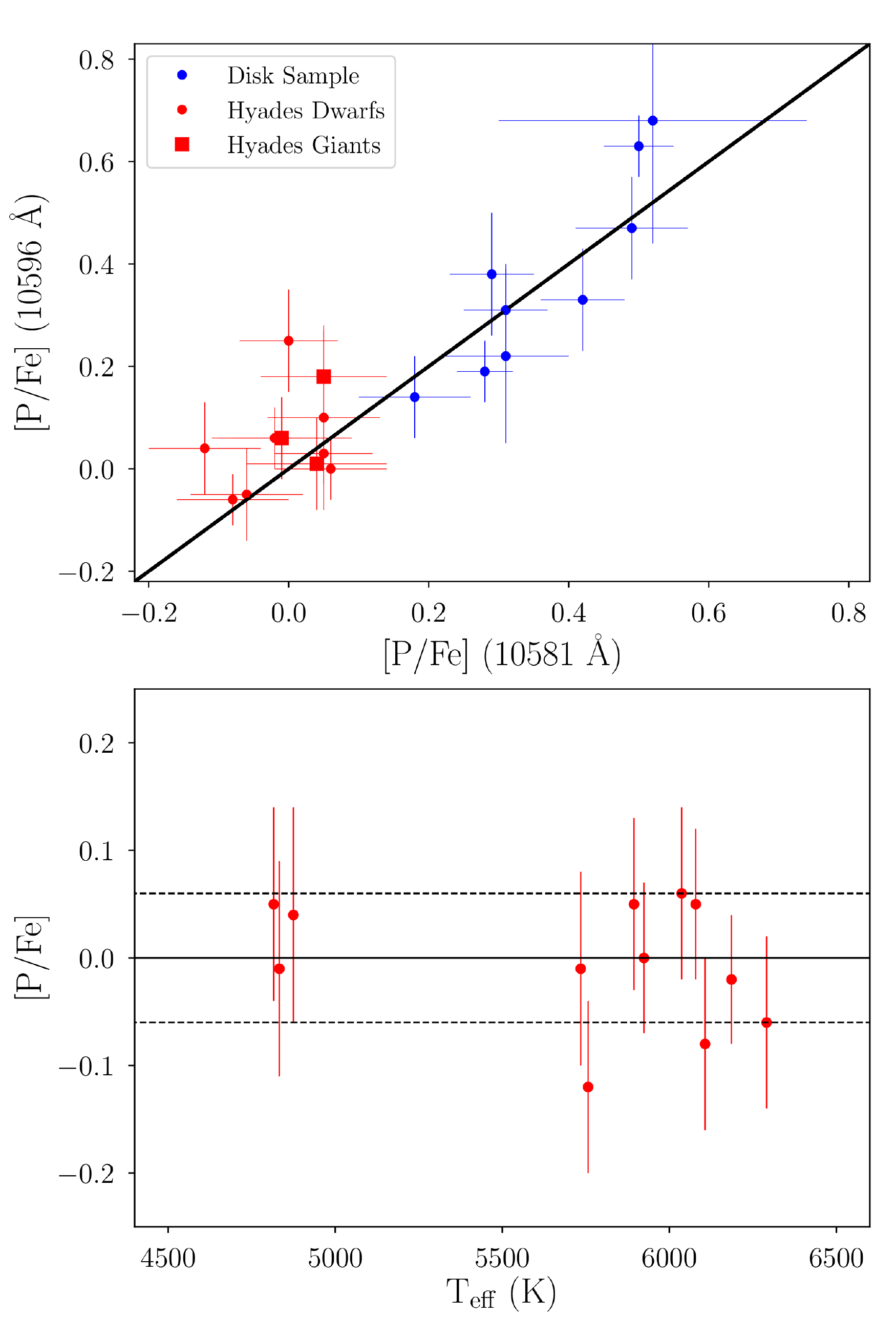}
	\caption{Blue points represent the disk sample of stars with atmospheric parameters from \citet{bensby14}, red points are Hyades dwarfs, and red squares are Hyades giants. Top Panel: Comparisons of P abundances derived from the 10581 $\mbox{\AA}$ P I lines to the 10596 $\mbox{\AA}$ line. Solid black line demonstrates a line with a slope of one. Bottom Panel: [P/Fe] ratio against effective temperature. Abundances from the 10581 $\mbox{\AA}$ are plotted. Solid lines represents the cluster mean [P/Fe] ratio of 0.00 and dashed lines represent the standard deviation of  $\sigma _{[P/Fe]}$ = 0.06 dex. \label{fig:line_comp} }
	\end{figure}

We took the average abundance for the 9 Hyades dwarfs in our sample, which span in 5735 K  $<$ T$_{\mathrm{eff}}$ $<$ 6291 K and compared that to the average of the three Hyades giants (T$_{\mathrm{eff}}$ $\sim$ 4800 K). We found the average abundance of the dwarf stars is --0.01 $\pm$ 0.06 dex and for the giants is 0.03 $\pm$ 0.03 dex (standard deviation for the sample of stars is quoted as the uncertainty). Additionally, the abundances for each P I feature versus T$_{\mathrm{eff}}$ is plotted in the bottom panel of Fig. \ref{fig:line_comp}. For both lines, no significant dependence between effective temperature and abundance is found. 

\section{Discussion}

\label{sec::discussion}

\subsection{Galactic Chemical Evolution of Phosphorus in the Disk}

\begin{figure*}[t!]
	\centering 
 	\includegraphics[trim=0cm 0cm 0cm 0cm, scale=.72]{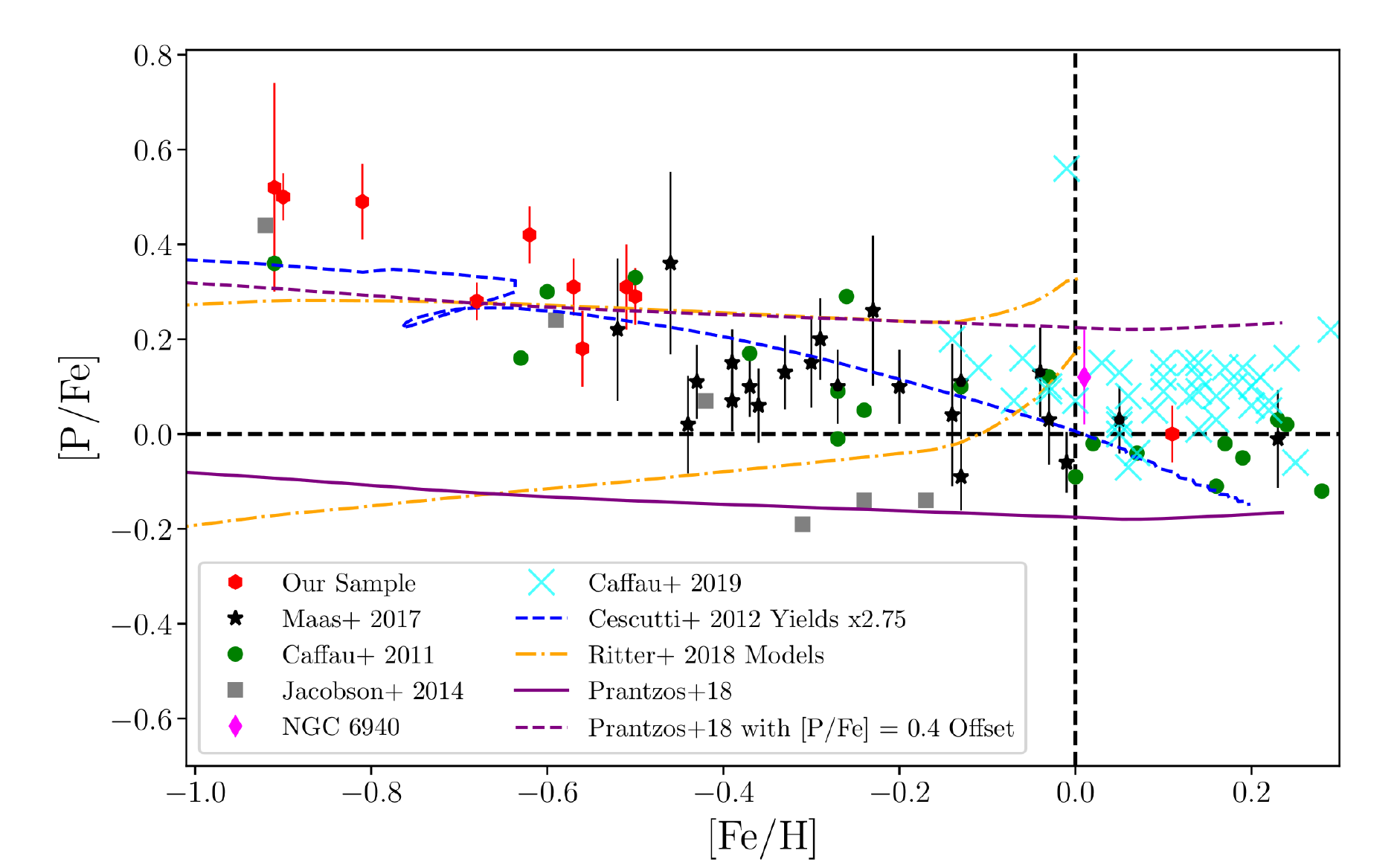}
	\caption{[P/Fe] ratios from this sample (red octogons), \citet{maas17} (black stars), \citet{caffau11} (green circles), \citet{jacobson14} (grey squares), \citet{caffau19} (cyan crosses) compared to chemical evolution models. The blue dashed line is a model with the yields increased by a factor of 2.75, both from \citet{cescutti12}. The orange dashed lines are chemical evolution models from \citet{ritter18} with O-C shell merger percentages of 0$\%$ (model with lowest [P/Fe] ratios predicted) and 10$\%$ model. Chemical evolution model from \citep{prantzos18} is shown as a purple solid line and the model arbitrarily increased by [P/Fe] = 0.4 dex is shown as a purple dashed line. Dashed black lines indicate solar abundances. Red octogon at [Fe/H] = 0.11 represents the average [P/Fe] for the Hyades and the uncertainties represent the standard deviation of the abundance measurements of stars in the cluster. Magenta diamond represents [P/Fe] cluster average for NGC 6940 \citep{topcu19} \label{fig:p_fe_chem_evolution} }
	\end{figure*}

To augment previous studies on the chemical evolution of P, we have compared our [P/Fe] results to models from \citet{cescutti12,ritter18,prantzos18} along with previous P abundance measurements, shown in Fig. \ref{fig:p_fe_chem_evolution}. Our measurements in disk stars from --1 $<$ [Fe/H] $<$ --0.5 in particular help distinguish between chemical evolution models. The observed [P/Fe] ratios from our a chemical evolution model with yields increased by 2.75 from \citet{cescutti12} (original yields from \citealt{kobayashi11}) fit the results from our new sample with [Fe/H] $<$ --0.5 as well as previous P abundance measurements. The difference between the model and stellar abundances therefore follows the same metallicity dependence as the predicted chemical evolution model (the model slope), which constrains the potential mechanisms needed to enhance P abundances in the chemical evolution models. Possibly the neutron capture cross-section may need to be revisited or \citet{caffau11} suggested proton capture on Si may need to be considered as a possible P production pathway. 

We also compare the chemical evolution model of \citet{prantzos18} to our abundances in Fig \ref{fig:p_fe_chem_evolution}. We find the chemical evolution model under-predicts [P/Fe] abundances over our metallicity range. This result is consistent with the result obtained by \citet{cescutti12} models that adopted the \citet{kobayashi11} yields for massive stars without applying any empirical enhancement. When arbitrarily enhanced by a factor of +0.40 dex, we still find the slope under predicts the abundances near solar metallicity and under-predicts [P/Fe] ratios for stars with [Fe/H] $\sim$ --1. We note that adopting yields from stellar model with higher rotation velocity from the set of yields by \citet{limongi18} \citep[on which the][is based]{prantzos18} will possibly improve the comparison with the data of [P/Fe]. However, this will degrade  the fit obtained for many other elements, in particular nitrogen, as described in \citet{prantzos18}.

Another chemical evolution study by \citet{ritter18} found that varying the degree the carbon and oxygen shells in massive stars interact or merge increases the yields of the odd-Z elements, such as P. Four chemical evolution models were created by \citet{ritter18} with varying C-O shell merger percentages of 0$\%$, 10$\%$, 50$\%$, and 100$\%$. These models are compared to our data in Fig. \ref{fig:p_fe_chem_evolution}. Our abundances also demonstrate the model with no yield enhancement (0$\%$) predicts a [P/Fe] evolution too low for the measured abundances. The closest matching model has an enhancement of 10$\%$, however the model under-predicts [P/Fe] at the lowest metallicities ([Fe/H] $\sim$ --1.0) and over-predicts the [P/Fe] ratio near the solar [Fe/H] abundance. The C-O shell mergers seem able to produce an enhancement in the P production and this is indeed an interesting feature. Nevertheless, the overall trend obtained by the chemical evolution model appears to be inconsistent with the observational [P/Fe] to [Fe/H] relationship. Possibly, a variable ratio of C-O shell mergers events as a function of metallicity  could reconcile the model with observation, but similar to rotation for the \citet{prantzos18} model, this will produce  consequences in the prediction of other elements \citep[as scandium, see][]{ritter18}. To summarise, there are  hints from the newest theoretical yields indicating that a solution for the phosphorus production problem in massive stars is close, however currently an empirical modification to adopted yields is needed to fit the observational data, as the one adopted in \citet{cescutti12} starting from the \citet{kobayashi11} yields.

We also note inconsistencies at the high metallicity end, in particular our measurement of the Hyades open cluster and the previous abundance measurements from \citet{caffau11, maas17,caffau19} lie above the chemical evolution model. Adjustments to the chemical evolution model, an additional production may be needed, or adjustments to the metallicity dependence of supernovae yields may be needed to match [P/Fe] ratios at solar metallicity, as shown in Fig. \ref{fig:p_fe_chem_evolution}. One possible solution is P production from a metallicity dependent source that only produces enough P to be significant at higher metallicities. \citet{topcu19} found a star (MMU 152) in NGC 6940 that exhibited abundance patterns consistent with mass transfer from an AGB star companion (e.g. high N, low C) with a slight P enhancement of $\sim$ 0.19 dex relative to the mean value for the other cluster members. However, AGB models do not predict P abundance enhancement \citep{karakas16}. P abundance measurements in more metal rich stars or s-process enriched stars may demonstrate whether AGB stars play any role in P production at the high metallicity range.

\subsection{Nucleosynthesis of P; Comparisons to Additional Elements}

\begin{figure*}[t!]
	\centering 
 	\includegraphics[trim=0cm 0cm 0cm 0cm, scale=.62]{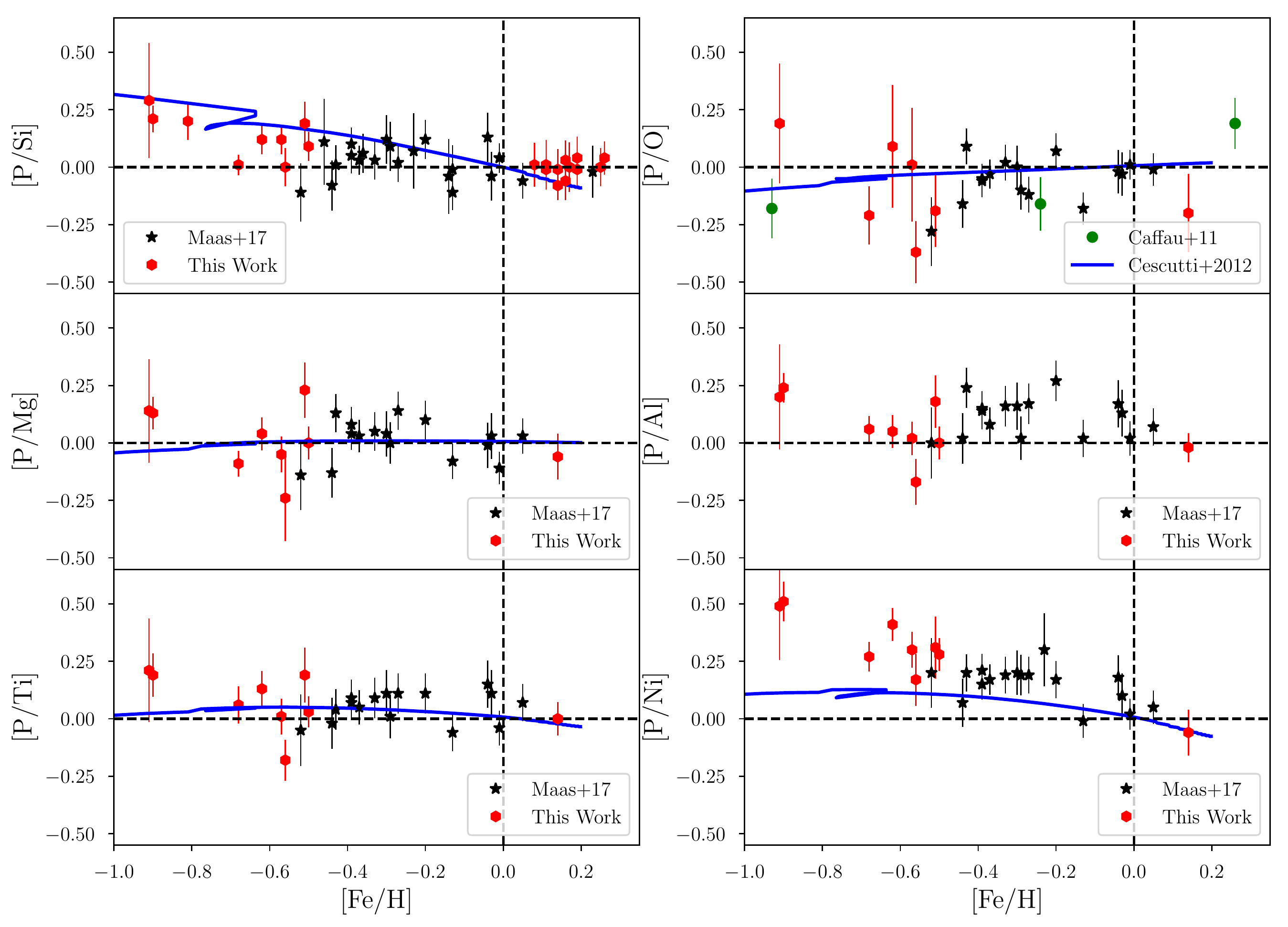}
	\caption{Black stars are measurements from \citet{maas17}, red hexagons from our sample, and green points from \citet{caffau11}. Dashed lines represent solar abundances. [P/X] versus [Fe/H] compared to the chemical evolution models (blue lines). \label{fig:p_o_si_comparison} }
	\end{figure*}

To further explore the nucleosynthesis of P, we compare our P abundances to $\alpha$ elements  (O, Mg, Si, and Ti), iron peak (Ni), and odd-Z elements (Al). For the $\alpha$ elements, O and Mg are thought to made almost entirely in hydrostatic burning in massive stars while Si and Ti are predicted to be made significantly in explosive burning in CCSNe with some contributions from Type Ia SN \citep{woosley95,travaglio04}. To make these comparisons, we use P and Si abundances in stars from \citet{maas17} and this work. Oxygen abundances are adopted from \citet{bensby14} for our stars and \citet{ramirez13} for the \citet{maas17} stars. Additionally, three stars from \citet{caffau11} had corresponding oxygen abundances derived by \citet{ramirez13}. We also adopt Mg, Al, Ti, and Ni abundances from \citet{reddy03} for the \citet{maas17} stars and from \citet{bensby14} for the current sample. Hyades oxygen abundances are obtained from \citet{takeda13} while abundances for the other elements are obtained from \citet{carrera11}. Finally, abundances for Arcturus were adopted from \citet{ramirez11}. In all cases, the adopted abundance ratios were re-scaled using the solar abundance ratios of \citet{asplund09}. All abundance ratios are plotted in Fig. \ref{fig:p_o_si_comparison}. 

% In the other panels, [P/O], [P/Mg], [P/Al], [P/Ti], and [P/Ni] ratios are plotted.
From Fig. \ref{fig:p_o_si_comparison} we first note that at the highest metallicity range (--0.5 $<$ [Fe/H] $<$ 0.2) the [P/Si], [P/O], [P/Mg], and [P/Ti] ratios remain consistent with the solar ratio. From --1.0 $<$ [Fe/H] $<$ --0.5, the [P/Si] ratios begins to increase, consistent with a model used in \citet{maas17}, derived from the P model from \citet{cescutti12}, Fe yields from \citet{kobayashi06}, and the Si yields of \citet{woosley95,francois04}. The nucleosynthesis prescription adopted for the other elements O, Mg, and Ti is similar to the one adopted in \citet{francois04}. We note the authors assumed empirical modifications - in particular for Mg - to match the trends and the solar abundances of several chemical species \citep{francois04}. Small differences in the Fe yields adopted cause small offsets between the chemical evolution model and the solar abundance ratios. The largest differences was the [P/O] ratio which was offset by $\sim$ --0.11 dex to normalize the [P/O] ratio to 0 at [Fe/H] =  0. In any case it is quite interesting to find that the [P/Mg] and [P/Ti] are also consistent with chemical evolution models, as shown in Fig \ref{fig:p_o_si_comparison}. Also the [P/O] chemical evolution model is consistent with the observed abundance ratios although there is significant scatter in the observed ratios. Compared to the other observed $\alpha$ elements, oxygen abundances show a dispersion in the data that may be due to the literature [O/Fe] abundances. We also tested [P/Ni] and found similar behavior as [P/Fe], expected for the iron-peak elements. However, the chemical evolution model under-predicts the [P/Ni] at metallicities lower than [Fe/H] $<$ --0.5. Similar slight offset is noticed in Fig. \ref{fig:p_fe_chem_evolution} and may indicate metallicity dependent changes are needed for the empirical P yields. Overall, the nearly solar observed [P/$\alpha$] ratios and agreement with chemical evolution models suggest P production is produced in similar environments as the $\alpha$-elements over the metallicity range of --1 $<$ [Fe/H] $<$ 0. 

Other odd-Z elements, Na and Al, decrease toward lower metallicities (e.g. the [Na/Mg] and [Al/Mg] ratios in \citealt{gehren06}) as the production method of both relies on neutron density and therefore has strongly metallicity dependent yields. We tested the [P/Al] abundance and found a slight offset from the solar ratio but consistent [P/Al] ratios at high metallicity. The decrease in neutron density for Al, however, is most significantly observed beyond [Fe/H] $<$ --1 \citep{gehren06}. Additionally, the [P/S] ratio was observed to be constant for stars with [Fe/H] $\gtrsim$ --1.0 \citep{caffau11} consistent with our $\alpha$ element ratios. 

Finally, small or no P contributions from Type Ia SN are suggested by our [P/$\alpha$] ratios. While we note a small increase in [P/Si] toward low metallicity, [P/Mg] and [P/Ti] are consistent with the solar ratio over our metallicity range; possibly also [P/O] shows the same trend, although the values for this ratio are scattered, especially at low metallicities. This suggests that small quantities of P are produced in Type Ia SN, less than than the silicon produced by Type Ia SN, and compatible with the negligible production of magnesium and oxygen. Studies of multiple APOGEE stars have suggested P may be significantly produced in Type Ia SN \citep{weinberg19}. However, our [P/Mg] ratios in particular do not suggest Type Ia SN are a significant source of P production compared to CCSNe. Additional P abundance measurements in metal poor disk stars are necessary to confirm these trends. 

\subsection{Phosphorus Abundances with Age}

The stars chosen for this exercise span a large range of ages, from the Hyades at 625 $\pm$ 50 Myr \citep{perryman98} to metal-poor thin and thick disk stars from \citet{bensby14}. We determined ages for stars with known P abundances from \citet{caffau11,maas17,caffau19} and the stars in this sample. Combining the multiple studies with new measurements from Gaia allows us to track the evolution of P with stellar age. Ages were determined from comparisons of the fundamental parameters luminosity, T$_{\mathrm{eff}}$, and [Fe/H] to Yonsei-Yale isochrones \citep{yi01}. Ages were computed using the \texttt{q$^{2}$} Python package \citep{ramirez14}, which adopted a maximum likelihood method to match the observed stellar parameters to stellar isochrones. 

\begin{figure*}[t!]
	\centering 
 	\includegraphics[trim=0cm 0cm 0cm 0cm, scale=.38, clip=True]{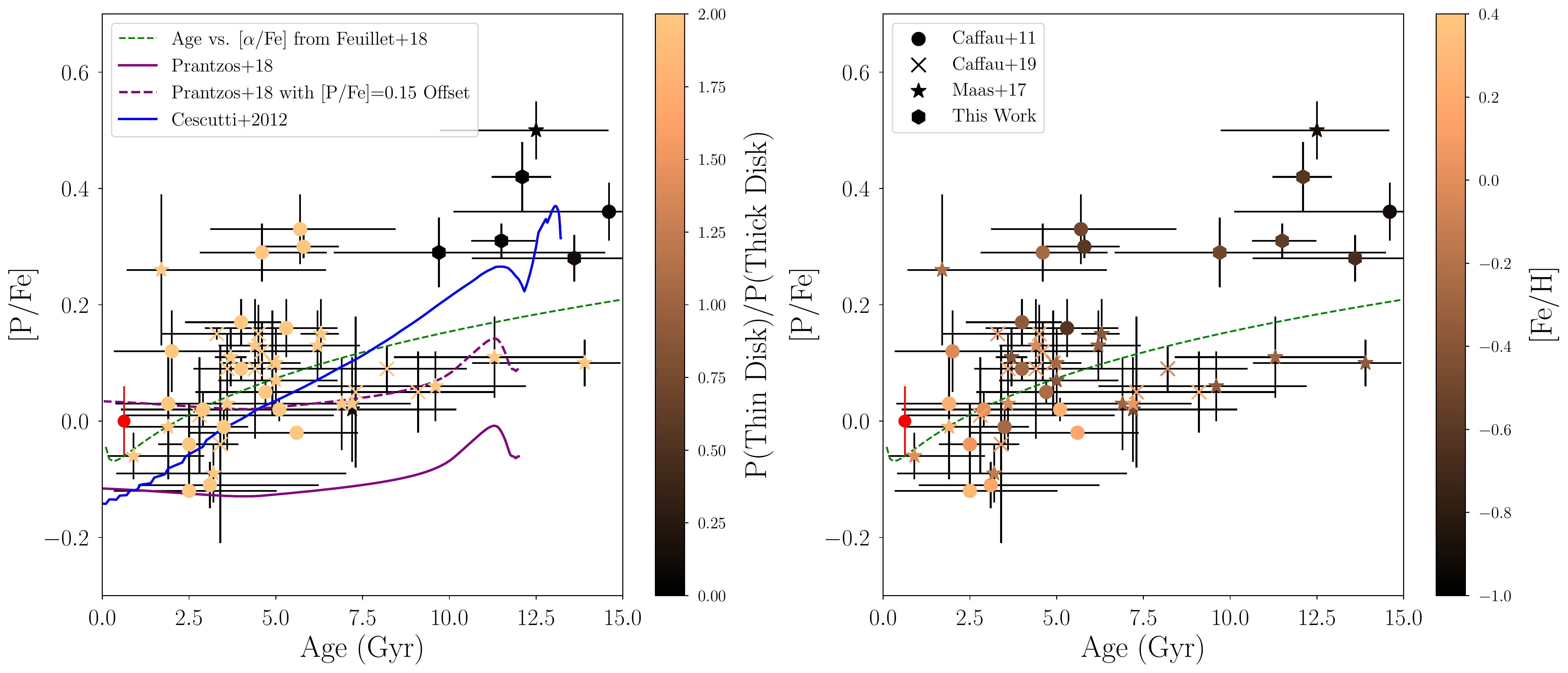}
	\caption{Ages and [P/Fe] ratios from this sample (octagons), \citet{maas17} (stars), \citet{caffau11} (circles), \citet{caffau19} (crosses). The red octagon represents the average Hyades abundance and the uncertainties represent the standard deviation of the abundance measurements of stars in the cluster. Green dashed line represents the fit to [$\alpha$/Fe] trend from \citet{feuillet18}. Left: Color bar shows the thin/thick disk probability ratio.  The blue line is a chemical evolution model from \citet{cescutti12} while the purple lines are models from \citet{prantzos18} Right: Color bar represents [Fe/H] for each star \label{fig:p_age}.}
	\end{figure*}

First we compiled parallaxes and V band photometry to measure the luminosity of each star in \citet{caffau11,maas17,caffau19}, and this work. Parallax measurements and their associated uncertainty from Gaia DR2 \citep{gaia18} were used to calculate the luminosity. We also adopted Johnson V band measurements from the Hipparcos/Tycho catalog \citep{esa97}, with an assumed uncertainty of 0.01 mags for each measurement. For a few stars, Gaia parallaxes or V band magnitudes were not measured. In those cases, V band measurements were adopted from the SIMBAD database and the parallaxes were available from the Hipparcos mission \citep{leeuwen07}. Effective temperatures, [Fe/H] abundances, and their uncertainties were adopted from the each atmospheric parameter source. In the case of \citet{caffau11} effective temperature and metallicity uncertainties were assumed at 100 K and $\delta$[Fe/H] of 0.15 dex. The parameters used and ages calculated are given in Table \ref{table::ages} and the results are plotted in Fig. \ref{fig:p_age}. Stars with 2$\sigma$ uncertainties larger than 5 Gyr were not included since these stars can not meaningfully constrain phosphorus production over time and distinguish between old populations ($\sim$ 10 Gyr) and young disk stars. 

We also determined thin and thick disk probabilities for the sample stars using phase space information available in the SIMBAD database. The UVW velocities were calculated using a python based UVW calculator\footnote{\url{https://github.com/dr-rodriguez/UVW_Calculator}}. The thin, thick disk, and halo membership probabilities based on the UVW velocities were calculated using the same methodology as \citet{ramirez13}. The UVW velocities and thin/thick disk membership probabilities are included in Table \ref{table::ages}. 

\begin{deluxetable*}{ c c c c c c c c c c c c c c }
%\tabletypesize{10pt} 
%\rotate
\tablewidth{0pt} 
\tabletypesize{\footnotesize}
\tablecaption{Star Age Estimate \label{table::ages}} 
 \tablehead{\colhead{HD Num} &  \colhead{[Fe/H]} &  \colhead{$\delta$[Fe/H] } & \colhead{Age} & \colhead{Age 2$\sigma$ lower} & \colhead{Age 2$\sigma$ upper} & \colhead{[P/Fe]} & \colhead{$\delta$[P/Fe]} & \colhead{[P/Fe] ref.}&\colhead{U} & \colhead{V} & \colhead{W} & \colhead{P(Thin)} & \colhead{P(Thick)} \\ \colhead{} & \colhead{(dex)} & \colhead{(dex)} & \colhead{(Gyr)} & \colhead{(Gyr)} & \colhead{(Gyr)} &  \colhead{(dex)} & \colhead{(dex)} & \colhead{} &\colhead{kms$^{-1}$}  & \colhead{kms$^{-1}$} & \colhead{kms$^{-1}$} & \colhead{} & \colhead{} } 
\startdata
HIP6949  & --0.50 & 0.05 &  9.7 &  6.7 & 14.5 &  0.29 & 0.06 & 2  &  43.73 & --101.21 &  34.74 & 0.00 & 0.96 \\
HIP7961  & --0.62 & 0.04 & 12.1 & 11.2 & 12.9 &  0.42 & 0.06 & 2  & --97.61 &  --76.08 &  34.87 & 0.02 & 0.94 \\
HIP10306 & --0.27 & 0.15 &  3.5 &  2.1 &  4.2 & -0.01 & 0.03 & 4  & --20.09 &  --12.22 &   4.43 & 0.99 & 0.01 \\
HIP15675 &  0.10 & 0.16 &  4.4 &  3.5 &  4.9 &  0.09 & 0.12 & 1  & --51.70 &  --44.23 &  13.49 & 0.90 & 0.10 \\
HIP17960 &  0.09 & 0.17 &  7.3 &  2.7 & 11.3 &  0.05 & 0.13 & 1  &  10.33 &  --58.95 & -14.74 & 0.78 & 0.22 \\
\enddata
\tablecomments{(This table is available in its entirety in machine-readable form.)}
\tablecomments{Ref: (1) \citet{caffau19}, (2) This Work,  (3) \citet{maas17}, (4) \citet{caffau11}}
\end{deluxetable*}

The compilation of [P/Fe], [Fe/H], thin and thick disk probabilities, and ages are plotted in Fig. \ref{fig:p_age}. A clear division in [P/Fe] exists between old (age $\gtrsim$ 10 Gyr) thick disk stars and the majority of the thin disk stars. Similar offsets  for the $\alpha$ elements are observed in older and younger stars (e.g. \citealt{bensby14, aguirre18}). The [P/Fe] ratio in the thin disk stars also appears to decrease slightly from a peak [P/Fe] ratio at $\sim$ 5 Gyr to the youngest stars at ~ 1 Gyr. However, more accurate ages are necessary to determine the evolution of [P/Fe] in the thin disk stars. 

We compare our [P/Fe] and ages to an empirical fit of [$\alpha$/Fe] vs. age in 712 red giant stars observed with APOGEE \citep{feuillet18}. The primary metallicity range of the \citet{feuillet18} sample is --0.5 $<$ [Fe/H] $<$ 0.4 and their $\alpha$ to age fit reproduced in Fig. \ref{fig:p_age} is consistent with [P/Fe] ratios measured at similar iron abundances. The few metal poor [P/Fe] measurements are offset at all ages from the empirical fit of \citet{feuillet18}. As a result, Fig. \ref{fig:p_age} provides additional evidence that P is made in CCSNe and correlates with $\alpha$ elements in stars with similar metallicities. 

We also find three high [P/Fe], low metallicity, young thin disk stars, HIP 40843, HIP 51523, and HIP 76716. These three stars are offset from other thin disk stars in Fig. \ref{fig:p_age}. The [P/Fe] abundances for these stars were determined by \citet{caffau11} and the ages for these stars are confirmed in other studies. HIP 40843 was found to also have an age of 3.39 Gyr, [Fe/H] = --0.26, and [O/Fe] = 0.05 from \citet{ramirez13}. HIP 51523 was determined to have an [Fe/H]= --0.48 and classified as a thin disk star by \citet{hinkel17}. Finally, \citet{holmberg09} measured [Fe/H] = -0.11 and an age of 3.1 Gyr for HIP 76716. Errors in age or disk membership cannot explain the high [P/Fe] abundances. \citet{caffau11} adopts [Fe/H] = --0.50 for HIP 40843, [Fe/H] = --0.60 for HIP 51523, and [Fe/H] = --0.26 for HIP 76716. Adopting the higher [Fe/H] ratios for these stars in the literature would lower the [P/Fe] ratio by $\sim$ 0.12 - 0.24 dex and would be consistent with other thin disk stars.

Finally, we compared our [P/Fe] ratios and ages to chemical evolution models. We find the model from \citet{cescutti12} (blue line in Fig. \ref{fig:p_age}) is in general agreement with the thin disk [P/Fe] ratios and thick disk abundances. The chemical evolution model from \citet{prantzos18} (purple lines in Fig. \ref{fig:p_age}) fits only the thin disk stars when enhanced by 0.15 dex. The agreement of the \citet{cescutti12} model with age provides additional evidence that the P yields must be offset without changing the metallicity dependence of the empirical yields. 

\section{Conclusions}
\label{sec::conclusions}
We measured P abundances in 21 stars; nine stars between --1.0 $<$ [Fe/H] $<$ --0.5 and 12 stars in the Hyades open cluster using the 10581 $\mbox{\AA}$ and 10596 $\mbox{\AA}$ P I absorption lines. The spectra were obtained with Phoenix on Gemini South and analyzed using MOOG spectral synthesis software. From our abundance measurements we have explored how P is produced in the Galaxy.
%\vspace{-.5cm}
\begin{enumerate}

\item{We found consistent average abundance in nine Hyades dwarf stars (--0.01 $\pm$ 0.06) and three Hyades giants (0.03 $\pm$ 0.03) dex. Temperature dependent systematic effects (e.g. NLTE effects) are therefore determined to be small for the P I features observed at this metallicity for FG dwarfs and giants.} 

\item{Our [P/Fe] ratios are consistent with the slope of chemical evolution models but offset over the metallicty range of --1 $<$ [Fe/H] $<$ --0.5. Models with P yields increased by a factor of 2.75 (by [P/Fe] $\sim$ 0.4 dex) match the chemical evolution models best. New models must therefore increase P production while keeping the same metallicity dependence over --1.0 $<$ [Fe/H] $<$ 0. Additionally, a small increase in P beyond [Fe/H] $\sim$ 0 is observed and may be the result of adjustments needed in SN metalliticy dependent yields and/or contributions from metallicity dependent P production (e.g. AGB stars) at high metallicities. }

\item{[P/O], [P/Mg], [P/Al], [P/Si], [P/Ti] and [P/Ni] ratios are compared to chemical evolution models. We find that P ratios to $\alpha$ elements are constant and approximately consistent with the solar ratio over nearly the entire --1.0 $<$ [Fe/H] $<$ 0.2 range, suggesting P is produced in the same environments as the $\alpha$ elements. We do note the [P/Si] ratio increases by $\sim$ 0.1 - 0.2 dex in the lowest metallicity stars as expected from models and suggests that Si is relatively more produced in Type Ia SN than P. Additionally there are hints that [P/O] decreases at our lowest metallicity range, however because [P/Mg] $\sim$ 0 over our entire metallicity range, we expect little to no contribution to P from Type Ia SN. }

\item{ The [P/Fe] trend with stellar age was compared to an empirical [$\alpha$/Fe] vs age relationship from \citet{feuillet18}. Our [P/Fe] ratios were consistent with the empirical fit over typical thin disk metallicities demonstrating P abundances are produced in CCSNe; specifically in the same environments as the $\alpha$ elements. Additionally, low metallicity, probable thick disk stars are offset to higher [P/Fe] ratios than the fit. Finally, our abundances are consistent with the [P/Fe] vs. age chemical evolution model from \citet{cescutti12}. }
\end{enumerate}

\section{Acknowledgements}
This work is based on observations obtained at the Gemini Observatory, which is operated by the Association of Universities for Research in Astronomy, Inc., under a cooperative agreement with the NSF on behalf of the Gemini partnership: the National Science Foundation (United States), the National Research Council (Canada), CONICYT (Chile), Ministerio de Ciencia, Tecnolog\'{i}a e Innovaci\'{o}n Productiva (Argentina), and Minist\'{e}rio da Ci\^{e}ncia, Tecnologia e Inova\c{c}\~{a}o (Brazil). The Gemini observations were done under proposal ID GS-2017B-Q-47. We thank Steve Margheim for his assistance with the Gemini South Telescope observing run. This research has made use of the NASA Astrophysics Data System Bibliographic Services, the Kurucz atomic line database operated by the Center for Astrophysics. We thank the anonymous referee for their thoughtful comments and suggestions on the manuscript. This research has made use of the SIMBAD database, operated at CDS, Strasbourg, France. This publication makes use of data products from the Two Micron All Sky Survey, which is a joint project of the University of Massachusetts and the Infrared Processing and Analysis Center/California Institute of Technology, funded by the National Aeronautics and Space Administration and the National Science Foundation. We thank Eric Ost for implementing the model atmosphere interpolation code. C. A. P. acknowledges the generosity of the Kirkwood Research Fund at Indiana University. Z. G. M. acknowledges the Indiana University College of Arts and Sciences for research support via a Dissertation Research Fellowship. G. Cescutti acknowledges financial support from the European Union Horizon 2020 research and innovation programme under the Marie Sk\l odowska-Curie grant agreement No. 664931 and from the EU COST Action CA16117 (ChETEC).

\software{\texttt{IRAF} \citep{tody86,tody93}, \texttt{MOOG} (v2017; \citealt{sneden73}),  \texttt{pymoogi}, \texttt{scipy} \citep{jones01}, \texttt{numpy} \citep{walt11}, \texttt{matplotlib} \citep{hunter07}, \texttt{emcee} \citep{foreman13}, \texttt{UVW calculator} (\url{https://github.com/dr-rodriguez/UVW_Calculator})}

\end{document}